\documentclass[prb,preprintnumbers,amsmath,amssymb,twocolumn]{revtex4}

\usepackage{graphicx}
\usepackage{dcolumn}
\usepackage{bm}

\begin{document}
\title{A Quantitative Solution to the Kondo Lattice Problem}

\author{Alex Breta\~{n}a$^{1,2}$, Sean Fayfar$^{1,3}$,  and Wouter Montfrooij$^1$} 
\affiliation{$^1$ Department of Physics and Astronomy, University of Missouri, Columbia, Missouri 65211, USA.\\
$^2$Advanced \& Energy Materials Group, Savannah River National Laboratory; SC 29802.\\
$^3$Nuclear Reactor Laboratory, Massachusetts Institute of Technology; Cambridge MA 02139.}
\begin{abstract}
{Metallic Kondo Lattice systems that have been prepared to exhibit a competition between ordering of magnetic moments and shielding of those moments by the conduction electrons down to absolute zero display unusual low-temperature responses. Here we show that the dominant response of such systems is caused by two quantum effects: zero-point motion of the ions, and the size of the system restricting the allowed wavelengths of fluctuations. This zero-point motion of the ions induces a broad distribution in Kondo shielding temperatures that renders the assumption of a uniform heavy-fermion ground state untenable. However,  letting go of this assumption and instead incorporating these two quantum effects leads to percolation physics that quantitatively captures the non-Fermi liquid response in both stoichiometric and doped quantum critical compounds, allowing for a unified description of all quantum critical systems.}
\end{abstract}
\maketitle
\section{Introduction}
Metallic systems that have magnetic ions embedded in their unit cells undergo a competition between Kondo shielding \cite{1} of the magnetic moments by the conduction electrons, and long-range ordering of those moments by the RKKY-interaction \cite{2} mediated by the conduction electrons. In such Kondo Lattice systems this competition can be manipulated so that it survives down to 0 K \cite{3,4,5}, referred to as the quantum critical point (QCP). The low temperature response of such systems, manifest in specific heat, susceptibility, resistivity, and neutron scattering measurements, shows a marked departure \cite{5} from Fermi-liquid theory for metals. This departure has been qualitatively ascribed \cite{4} to intrinsic differences between a zero Kelvin order-disorder transition and transitions that occur at finite temperatures. However, no uniform description has emerged.\\

Theoretical descriptions have been put forward to describe the unusual low-temperature response, ranging from a distribution of Kondo shielding temperatures for heavily-doped quantum critical systems to a wide variety of ideas for (almost) stoichiometric systems.  Examples of the latter are locally critical fluctuations \cite{local}, a spin-density wave instability of the Fermi surface \cite{sdw0,sdw}, as well as the onset of AF-ordering in the presence of strong ferromagnetic fluctuations \cite{20} driven by the breakup of heavy quasiparticles \cite{18}. Understandably, attention has focused on trying to understand the nearly stoichiometric systems as these are free from the obfuscating effects that a distribution of Kondo shielding temperatures brings along with it. However, these theoretical efforts are based on the flawed assumption that the ground state of a  heavy-fermion system is uniform on all length scales.\\

\begin{figure}[t]
\begin{center}
\includegraphics*[viewport=110 140 550 410,width=85mm,clip]{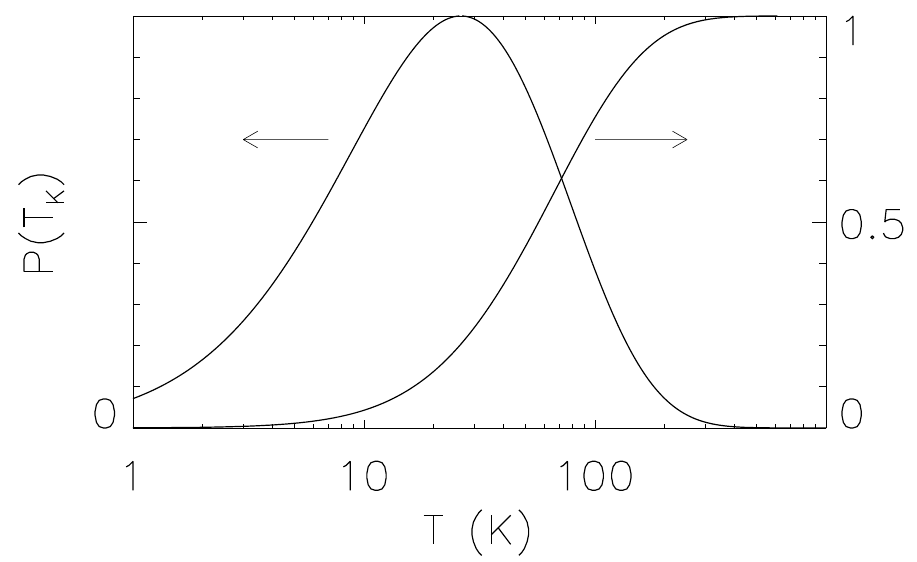}
\end{center}
\caption{Instantaneous distribution of Kondo shielding temperatures (left axis) and resulting fraction of ions where the formation of a magnetic singlet is favored (right axis) for URu$_2$Si$_2$. Note the logarithmic temperature axis. The average Kondo temperature was chosen to be 75 K, the conduction bandwidth was taken to be 0.4 eV based on theoretical estimates \cite{bandwidth}, and the Debye-Waller factor for the U-Ru distance was fixed at 0.006 nm based on experimental observations \cite{RuU}.} 
\label{figkondo}
\end{figure}

The ground state of a stoichiometric heavy-fermion system cannot be characterized by a well-defined Kondo energy scale that is uniform throughout the sample. In fact, the opposite is true as a simple calculation shows. Using the known separation r dependence of the hybridization strength \cite{6} between f- and d-ions, we can rewrite the standard expression\cite{4,6} for the Kondo energy scale $k_BT_K=De^{-1/J\rho(\epsilon_F)}$ as $k_BT_K=De^{-Ar^{12}}$ for f-d hybridization (with D the bandwidth, J the Kondo coupling strength, and $\rho(\epsilon_F)$ the density of states at the Fermi level). Differentiating this expression and eliminating the unknown constant A in favor of the Kondo energy scale we obtain:
\begin{equation}
\frac{dT_K}{T_K}=12 ln(D/k_BT_K)\frac{dr}{r}.
\end{equation}
Typical bandwiths values are of the order of 0.5 eV, typical values for $T_K$ are 20-30 K (yielding ln($D/k_BT_K) \sim$ 4-6), whereas typical values for dr/r due to zero-point motion are 1.5-2\% at zero kelvin, as measured in diffraction experiments. As such, typical d$T_K/T_K$ values are $\sim$ 50-100\%.  We show the distribution of shielding temperatures for URu$_2$Si$_2$ based on experimental values in Fig. \ref{figkondo} where we used a Gaussian distribution P(r) for the Debye-Waller factor with conservation of probability: P(T$_K$)dT$_K$=P(r)dr. It is clear from this figure that a mere 2\% distribution in equilibrium positions results, quite counterintuitively, in a distribution of large width in Kondo shielding temperatures. This unavoidable spread in Kondo energy scales gives rise to a highly non-uniform ground state at any instant in time, and provided the electronic time scales are much faster than the ionic ones, we can expect stoichiometric quantum critical systems to exhibit all the hallmarks of heavily-doped systems. \\

The only situations in which we can safely ignore this non-uniformity is under circumstances where the adiabatic approximation breaks down, or when we are looking at length scales much larger than the scale of the spontaneous zero-point motion induced variations of the Kondo scale. We are not aware of any recorded instance of the adiabatic approximation breaking down (as manifest in an unusual temperature dependence of phonon frequencies). The latter situation is analogous to that of the Universe: on very large length scales the Universe is uniform and isotropic, but on somewhat smaller length scales it is quite the opposite with galaxies and voids appearing. Given this intrinsic variation, it is unlikely that theories based on the assumption of uniformity will ever lead to a unified description of quantum critical phenomena.\\

We present a quantatative, unified description of doped and stoichiometric quantum critical systems. In recent work \cite{8,11} we have shown, by focusing on the distribution of Kondo energy scales, that we can not only understand the quantum critical behavior in heavily-doped  Ce(Ru$_{0.25}$Fe$_{0.75}$)$_2$Ge$_2$\cite{11} but that these distributions were also relevant \cite{8} to stoichiometric CeRu$_2$Si$_2$. In particular, the inconsistencies of observing fluctuating moments in field-dependent ac-susceptibility experiments that exceeded isolated, unshielded Ce-moment values were resolved \cite{8}. In here, we extend this analysis to three more quantum critical systems and show that this wide variety of doped and stoichiometric systems is dominated by the spread in Kondo energy scales and the basic rules of quantum mechanics. We make our case through a quantitative analysis of literature data for five heavily studied quantum critical systems (the stoichiometric systems YbRh$_2$Si$_2$,\cite{18,21} CeRu$_2$Si$_2$,\cite{8,12,25}  CeCu$_6$,\cite{23,24} and the heavily doped compounds UCu$_4$Pd\cite{15} and Ce(Ru$_{0.25}$Fe$_{0.75}$)$_2$Ge$_2$\cite{10,11} and show that their response can be reproduced by that of a collection of magnetic clusters, as detailed in the next section. In the process, we offer a much simplified interpretation of the rich phase diagram \cite{21} of YbRh$_2$Si$_2$.

\section{Theory}
The interaction strength between magnetic moments and conduction electrons depends exponentially sensitively on the separation between ions \cite{6}. Small changes in these separations, achieved by hydrostatic or chemical pressure (doping by smaller or larger ions) drive a system through the order-disorder transition \cite{3,7}. At the unit cell level, however, there is no fixed interatomic separation. Random doping leads to a static distribution of separations, while the intrinsic zero-point motion (ZPM) of ions also leads to a distribution of separations at any given moment in time. The two effects are of similar magnitude with departures from equilibrium lattice sites of the order of 0.004-0.006 nm at T= 0 K \cite{8}. The distribution of separations leads to a distribution of Kondo temperatures T$_K$, with the width of the distribution of the order of tens of Kelvin \cite{8} even for Debye-Waller factors as small as 0.003 nm. Since the time scale for electronic motion is much faster than that of ionic motion, there is very little difference from a conduction electron perspective between static and dynamic distributions. This is none other than the adiabatic approximation from phonon theory and it is the reason why the critical physics of both stoichiometric and doped quantum critical systems can be compared directly, such as in the parametric study of CeRu$_2$(Si$_x$Ge$_{1-x}$)$_2$ \cite{7}.\\

A distribution of T$_K$ leads to percolation physics \cite{9} upon cooling. Depending on the local T$_K$, a magnetic moment will be either still be present or it will be (effectively) Kondo shielded (in the sense that a magnetic singlet formation is energetically favored). With increased cooling, more and more moments become shielded, and groups of moments (clusters) emerge that are separated from other moments by shielded ones. Such clusters acquire special characteristics because of quantum mechanical finite-size effects: the RKKY-interaction \cite{10,11} will align the moments within such a cluster with their neighbors. The moments align because the energy cost is too high to keep them misaligned \cite{11}. The energy required to misalign moments is inversely proportional to the wavelength of the disordering fluctuation, and this wavelength cannot be arbitrarily large as it has to match the size of the cluster. At low T, there is not enough thermal energy available to prevent the moments on a cluster from aligning. Thus, $<J_ie^{i\vec{Q}.\vec{r}ij}J_j>\neq 0$ for two moments ($J_i$ and $J_j$) on the same cluster with separation $\vec{r}_{ij}$, and $\vec{Q}$ the RKKY ordering vector. Lastly, once these moments align, then the Kondo shielding mechanism is severely impeded \cite{11} as this mechanism involves a spin-flip exchange process \cite{1} that is impeded by the ordered local environment. This impediment results in a change of universality class \cite{16} compared to the regular\cite{9} critical exponents for percolation in such a way that the new set of exponents violates\cite{sean} the Harris criterion \cite{harris}.\\
 
\begin{figure}[t]
\begin{center}
\includegraphics*[viewport=0 0 730 500,width=85mm,clip]{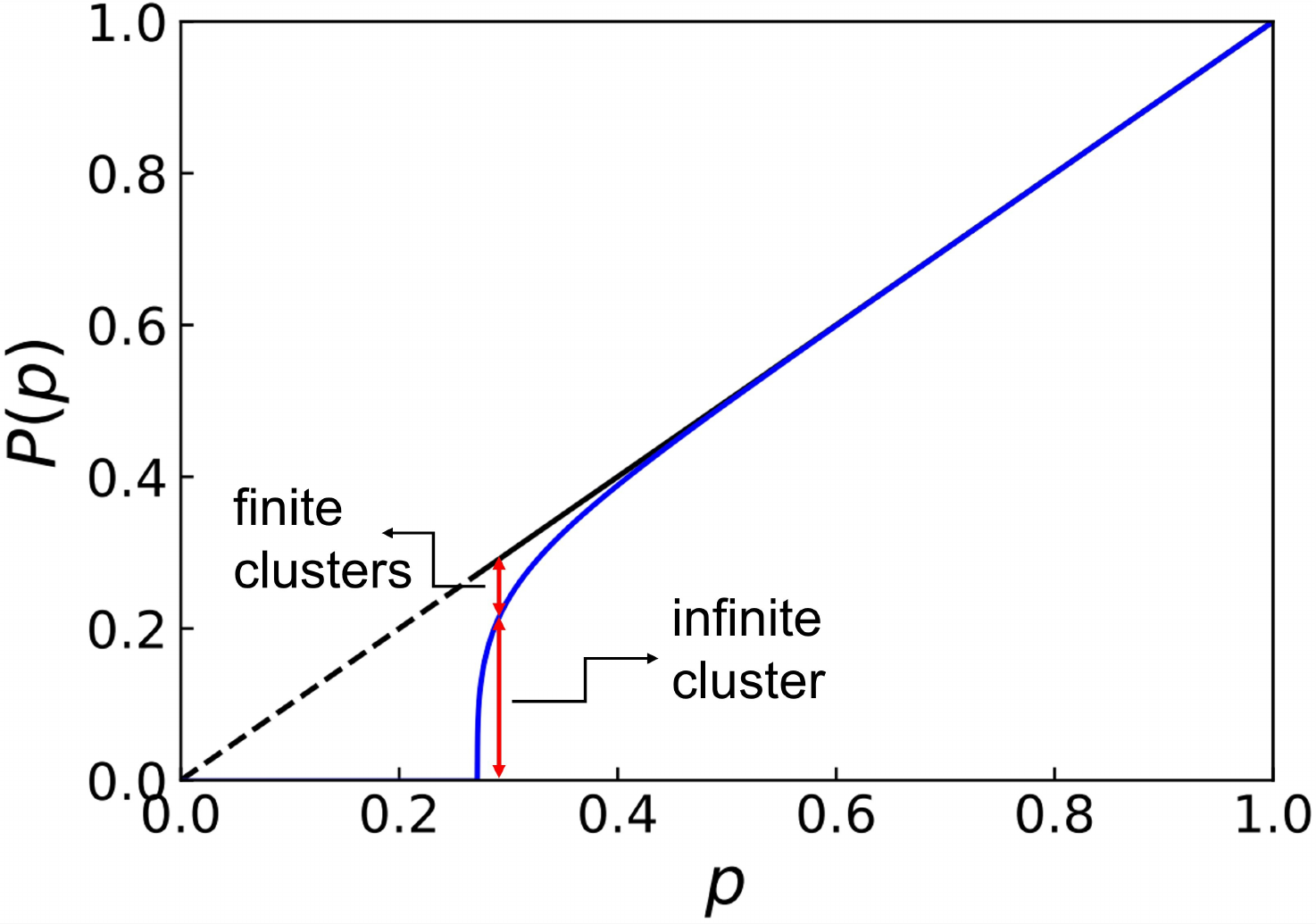}
\end{center}
\caption{(Color online) The strength of the infinite cluster P as a function of occupancy p for a body-centered structure. At high occupancies, this lattice spanning cluster follows the occppancy, but near the percolation threshold p$_c$= 0.272 \cite{16} it follows a power law in p-p$_c$. Equating the strength of P to S(T)/Rln2 yields the occupancy (number of unshielded moments) at any given temperature. The distance between the curves p=p (line through origin) and P(p) (solid curve) represents the mass locked up in the finite clusters.} 
\label{cluster}
\end{figure}

Upon cooling, increasingly more moments become shielded, with increasingly larger clusters forming. The shielding of these moments, and the forced ordering of the moments within isolated clusters is reflected in the specific heat curve that follows the loss of magnetic entropy. All the magnetic entropy of the system is contained in the lattice spanning (infinite) cluster: the collection of unshielded moments (Fig. \ref{cluster}) that have not broken apart into isolated clusters and therefore, are not forced to align with their neighbors \cite{11}. The susceptibility follows a similar pattern: while the moments on the infinite cluster remain free to respond to a magnetic field, the shielded moments no longer contribute, and the moments on isolated clusters now behave as a superspin consisting of the overall net moment of all the ordered moments on the cluster. In addition, some of the superspins can be so large that they easily align with and amplify small external fields, giving rise to a super-paramagnetic response \cite{8} that can easily be mistaken for ferromagnetic fluctuations.\\

In neutron scattering experiments \cite{10,11,12} the emergence of clusters is observed as a unique signature localized around the RKKY-ordering wave vector whose intensity increases upon cooling but with finite width, reflecting the absence of long-range order. Moreover, since all the moments in finite clusters have lined up, these widths are independent of lattice direction \cite{8,10} when counted in lattice units, even for non-cubic systems. We verified the above in both CeRu$_2$Si$_2$ \cite{8} and Ce(Ru$_{0.25}$Fe$_{0.75}$)$_2$Ge$_2$ \cite{10,11}, providing direct evidence for the emergence of clusters in these systems. Also, the large discrepancies in Ce moment values in CeRu$_2$Si$_2$ (10$^{-4}$ – 4 $\mu_B$) inferred from different probes were largely resolved \cite{8} by modeling the response at low T by the response of a collection of clusters near the percolation threshold  p$_c$. \\

We demonstrate in this paper that the cluster scenario can be tested even in systems that have inconclusive neutron scattering results owing to a lack of accuracy (YbRh$_2$Si$_2$) \cite{13}, competing ordering wave vectors (CeCu$_6$) \cite{14}, or a cubic crystal structure (UCu$_4$Pd) \cite{15}, by comparing experiments to the predictions of percolation computer simulations \cite{16} as well as via a direct comparison between the specific heat and susceptibility that is implicit in the cluster scenario and detailed below. \\

When clusters form, their constituent moments align with their neighbors and their magnetic entropy is lost, only leaving their superspin orientation degree of freedom. Thus, in the absence of a magnetic field B, all recoverable entropy is contained in the unshielded moments that form the infinite cluster \cite{9}. As such, the entropy S of the system is a direct measure for how many moments are part of the lattice spanning (infinite) cluster (Fig. \ref{cluster}), and provided these moments barely interact with each other (at high T), we expect \cite{8} its susceptibility $\chi_{\infty}$ to be given by a product of how many such unordered, unshielded moments exist and the free susceptibility of an uncorrelated effective moment of strength $\mu_{eff}$:
\begin{equation}
\chi_{\infty}(B,T) = \frac{S}{R ln 2}\chi_{free}(B,T,\mu_{eff}) \approx  \frac{S}{R ln 2} \frac{\mu_{eff}^2}{k_BT}.
\label{connection}
\end{equation}
In here, we assumed a ground state doublet (Rln2 entropy), while the last part of the equation is only valid for a hard-axis system in the low B/T limit. At high T, there will be few finite clusters and we expect $\chi_{\infty}$ to be close to the experimental susceptibility $\chi_{exp}$.\\

At low temperatures, the superspins of the collection of isolated clusters will react to an external magnetic field. In order to quantify their contribution to the susceptibility, we use the results of percolation computer simulations performed at the percolation threshold p$_c$. The details of the simulation are given elsewhere \cite{8}, yielding the following characteristics: assuming anti-ferromagnetic (AF) order, at p$_c$ there are 0.0078 clusters with a non-zero superspin per lattice site, and on average, 1.87 uncompensated moments per cluster (0.0147 moments per site). The full distribution is shown in Fig. 8 of reference \cite{8}.\\

Comparing measurements to the susceptibility of the simulated cluster distribution is not expected to yield exact agreement for a few reasons, but nonetheless serves as indirect proof for the presence of clusters. First, a moment value needs to be assigned to the uncompensated moments that form a cluster; we use $\mu_{eff}$ from Eq. \ref{connection} for this. Second, the distribution at p$_c$ is an overestimate of the actual one as the system will not be exactly at the threshold since raising T drives it away from p$_c$. We adjust for this by scaling the response to the total mass of the cluster distribution as graphically sketched in Fig. \ref{cluster}. Third, in the case where the actual ordering wave vector deviates from AF-order, the distribution yields an underestimate because the superspins in the simulation were calculated assuming AF order \cite{8}.  This distinction is most relevant for the smallest clusters. In order to compensate for this, we have run simulations to evaluate the expected average cluster moment for small-sized clusters based on the measured RKKY-ordering wave vectors. We then replace the superspins of the smaller clusters in the simulations with the newly evaluated cluster moment. This procedure is not exact as there exists a spread in cluster moments for clusters of fixed size depending on the morphoplogy of a particular cluster. Notwithstanding these drawbacks, we find that we obtain a quantatative agreement between experiments and the predictions of the cluster scenario using $\mu_{eff}$ as the only adjustable parameter, as shown in the results section.\\

We note that $\mu_{eff}$ can only be adjusted within a tight margin and as such is not a completely free parameter. Its value cannot exceed the free moment value, but it has to exceed the value estimated from low-temperature measurements as these are based on the assumption that all ions contribute equally, rather than a fraction of them being shielded. Moreover, $\mu_{eff}$ can be directly measured in high-field neutron scattering experiments \cite{13}  by assessing the energy associated with intra-doublet transitions. For example, our analysis (presented in the next section)  for YbRh$_2$Si$_2$ infers a moment  of $\mu_{eff}$= 1.74 $\mu_B$, with bulk estimates \cite{20} putting it at 1.4 $\mu_B$ and measured intra doublet transitions \cite{13} at (1.9 $\pm$ 0.1) $\mu_B$.\\

In addition to the direct observation of clusters by means of correlation length determination using neutron scattering, and the inference through specific heat and susceptibility data, there is a method applicable to YbRh$_2$Si$_2$ to infer the existence of clusters at low temperatures. We do not know whether this method extends to other quantum critical systems as well. At the lowest temperatures, and in the absence of clusters, the Fermi-liquid low-temperature magnetization and specific heat are both proportional to the density of states at the Fermi-level g($\epsilon_F$). This density depends on the magnetic field B. In YbRh$_2$Si$_2$ it was found \cite{18} that the low-T specific heat depended on field B as a power law, implying that g($\epsilon_F$) varied as a power-law g($\epsilon_F$)$\sim 1/B^{\alpha}$. Using the standard expression for the low-temperature electronic magnetization M=$\mu$Bg($\epsilon_F$), we can relate the uniform (dc) and ac-susceptibilities as follows:

\begin{equation}
\chi_{dc}=M/B=\mu g( \epsilon _F); \chi_{ac}=\frac{dM}{dB}=(1-\alpha)\chi_{dc}.
\end{equation}
Thus, when the low-T field dependence of the specific heat is given by a power law, we expect $\chi_{dc}$ and $\chi_{ac}$ to differ by a factor of (1-$\alpha$). However, should clusters with superspins be present, then they will differ more. At the lowest temperatures, under conditions of B/T $>>$ 1, the clusters will not show up in the ac-susceptibility as they have fully aligned with the primary magnetic field, but they will show up in the magnetization. In particular, the following expression can be used for verification of the presence of clusters at low temperatures, as well as for an estimate of the average superspin of the collection of clusters:
\begin{equation}
\mu_{excess}=[\chi_{dc}-\chi_{ac}/(1-\alpha)]B
\label{acdc}
\end{equation}
In the absence of clusters, $\mu_{excess}$ should be equal to zero, in their presence it should be a non-zero value independent of the strength of magnetic field. In the next section we apply this procedure to the susceptibility data for  YbRh$_2$Si$_2$ .

\section{Results and Discussion}
\begin{figure}[t]
\begin{center}
\includegraphics*[viewport=40 110 545 510,width=85mm,clip]{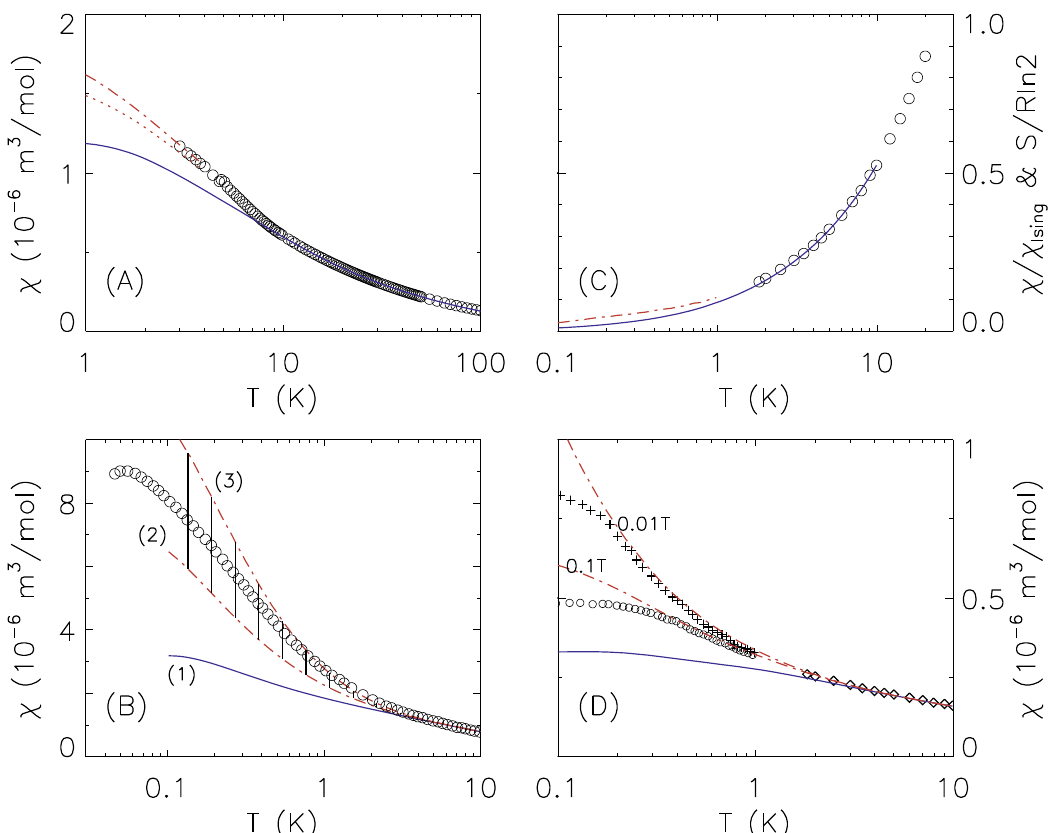}
\end{center}
\caption{(Color online) Comparison of the measured uniform susceptibility with the prediction of Eq. 1. (A) The uniform susceptibility (symbols) for Ce(Ru$_{0.25}$Fe$_{0.75}$)$_2$Ge$_2$ and the entropy (solid blue curve) \cite{11} scaled using Eq. 1 with $\mu_{eff}$= 1.43 $\mu_B$ for best agreement. The dotted curve is the susceptibility for the simulated collection of clusters at the quantum critical point for this $\mu_{eff}$ value added to the prediction of Eq. 1. for AF-ordered clusters, the dashed dotted curve was calculated by replacing the moments of the five smallest clusters by the average moment for clusters with the measured ordering wave vector Q=(0,0,0.45) \cite{10}. (B) Same as (A) but for YbRh$_2$Si$_2$ \cite{19} for B= 0.025 T (symbols). Curve (1) shows the scaled entropy (using $\mu_{eff}$ = 1.74 $\mu_B$), curve (2) is the contribution of the AF collection of clusters added to Eq. 1, curve (3) is the same with the five smallest cluster moments replaced  by the average moment for clusters with ordering wave vector Q=(0.14,0.14,0) \cite{13}. (C) High-T susceptibility (symbols) for UCu$_4$Pd measured in a field of 0.5 T \cite{15}, and low-T in a field of 0.01 T (dashed curve, T $<$ 1 K)\cite{17}. $\chi$  has been scaled to that of an Ising system ($\mu_{eff}$ = 0.8 $\mu_B$). The measured entropy \cite{15} is given by the solid curve. (D) Susceptibility \cite{15,17} of UCu$_4$Pd at 0.5 T (diamonds), 0.1 T (circles), and 0.01 T (+ symbols). The solid curve is the prediction of Eq. 1 ($\mu_{eff}$ = 0.8 $\mu_B$), with the cluster contribution added at the various fields (dashed lines). For the latter we adjusted the cluster mass for this AF face-centered cubic system from the AF-BCC simulations by the ratio of the respective percolation thresholds which is a measure of the surviving moments locked up in isolated clusters at p$_c$).} 
\label{fig1}
\end{figure}

We show the comparison between the uniform susceptibility $\chi_{exp}$ and S using Eq. (\ref{connection}) in Fig. \ref{fig1} (with $\mu_{eff}$ as an adjustable parameter) for the doped systems Ce(Ru$_{0.25}$Fe$_{0.75}$)$_2$Ge$_2$ \cite{11} and UCu$_4$Pd \cite{15,17},  and stoichiometric YbRh$_2$Si$_2$ \cite{18,19,20,21}. It is clear from this figure that $\chi_{exp}$ and S track each other precisely at the higher temperatures (T $>$ 5 K), with $\chi_{exp}$ exceeding the predicted value at low T once the entropy falls below ~0.4Rln2. The precise tracking is most easily seen in Fig. \ref{fig1}A and C (symbols and solid line through the points). Note that S= 0.4Rln2 is roughly the value where clusters first appear \cite{16} (the point where the curves p=p and P(p) start to deviate in Fig. \ref{cluster}) with their superspins starting to contribute to the susceptibility.\\

In order to compare the experimental results to those of the cluster scenario at all temperatures, we quantify the susceptibility of the isolated clusters by calculating the response of a simulated collection of clusters at p$_c$, as discussed in the preceding section. We show the results in Fig. \ref{fig1} for both anti-ferromagnetic order between the moments, as well as for RKKY-order by replacing the superspin of the smallest clusters by the average superspin of an ensemble of RKKY-ordered clusters of identical size. It can be seen in Fig. \ref{fig1} that the experimental results and the simulated ones (using the  $\mu_{eff}$ values determined for T $>$ 5 K) are in good qualitative agreement, demonstrating that the observed low-T response is that of a collection of clusters.\\

\begin{figure}[t]
\begin{center}
\includegraphics*[viewport=155 107 575 444,width=80mm,clip]{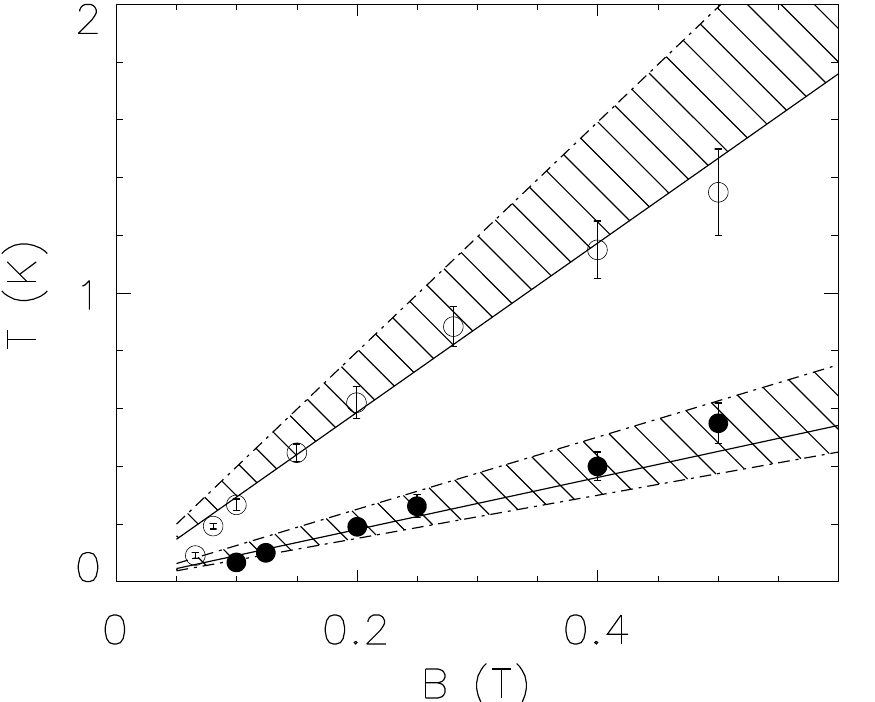}
\end{center}
\caption{The field dependence of the maxima in c/T (closed circles) and ac-susceptibility (open circles) \cite{19,21}.  The lines denote the maxima for a simulated distribution of clusters at p$_c$ for a ground state doublet ($\mu_{eff}$= 1.74 $\mu_B$), with the hatched areas indicative of the sensitivity of the predicted maxima to the ordering wave vector (dotted curve: AF ordering; solid curves: replacing the AF-zero moment of clusters of size 2 with the average moment calculated for Q=(0.14,0.14,0) \cite{13}; dashed-dotetd curves: replacing clusters up to size five with the average moment for ordering with  Q=(0.14,0.14,0).} 
\label{fig2}
\end{figure}

The presence of clusters significantly simplifies the interpretation \cite{21} of the phase diagram of YbRh$_2$Si$_2$. YbRh$_2$Si$_2$ is a heavy-fermion system close to the QCP \cite{18}, which can be reached by applying a modest magnetic field perpendicular to the hard axis (0.06 T) to inhibit long-range order at 0.07 K, or by 5\% doping of Ge for Si. Specific heat, resistivity, ac- and dc-susceptibility measurements have been performed as a function of temperature, magnetic field and chemical doping \cite{18,19,20,21}. The literature interpretation is that the QCP is characterized by the onset of AF-ordering in the presence of strong ferromagnetic fluctuations \cite{20} driven by the breakup of heavy quasiparticles \cite{18}, whereas two critical magnetic fields have been identified between which the system enters \cite{21} a non-Fermi liquid phase (with long range order at lower fields and Fermi-liquid behavior at higher fields). The richness of this phase diagram is reproduced in Fig. \ref{fig2}. Neutron scattering data revealed incommensurate short-range magnetic correlations \cite{13} [Q= (0.14,0.14,0)] as well as inelastic intra-doublet transitions ($\Delta E= 2\mu_{eff}B$) corresponding to $\mu_{eff}$ = (1.9$\pm$  0.1)$\mu_B$. Unfortunately, the neutron data were not accurate enough to ascertain whether the extent of the magnetic correlations is identical in all directions, or not.\\

The ordered phase in YbRh$_2$Si$_2$ (T$<$ 0.07 K, B$<$ 0.06 T) corresponds to a loss in magnetic entropy of 0.008 Rln2 and an ordered moment $\mu$$<$ 0.1 $\mu_B$ per Yb ion \cite{19}. In our cluster scenario, when the low temperature state of a system is characterized by a distribution of clusters with a superspin, then spontaneous order can arise between the superspins at low enough temperatures. From our computer simulations, the expected loss of entropy associated with an ordering of the superspins is 0.0078 Rln2, and the moment per Yb ion in the ordered state is given by 0.0147 $\mu_{eff}$, well below 0.1 $\mu_B$ per Yb ion. Therefore, we interpret the ordered state as a collection of magnetic clusters with their superspins aligned. \\

The presence of clusters also accounts for the non-Fermi liquid response observed in the specific heat and ac-susceptibility. Different peak values for these curves (Fig. \ref{fig2}) are expected for superspins: when the ground state is a doublet, c/T of a single cluster peaks at $\mu_{cluster}$B= 1.622 k$_B$T, while the ac-susceptibility for a two-level system peaks at $\mu_{cluster}$B= 0.772 k$_B$T. These peak positions were evaluated numerically for a two-level system displaying a gap $\Delta$= 2$\mu$B. Adding in the full (B,T)-dependence and averaging over all clusters in the simulations, we arrive at the peak positions shown in Fig. \ref{fig2}. As the exact peak positions are somewhat dependent on the net moments of the smallest clusters, we have presented our results as a bandwidth. It is clear, however, that the two different peak values for c/T and $\chi$ originate from one underlying mechanism: the presence of clusters. As an aside, even isolated moments would give rise to different peak values in c/T and the ac-susceptibility with a ratio of 1.622/0.772= 2.1, and as such, one should not automatically interpret the appearance of two different peak values as an indication of novel physics.\\

\begin{figure}[t]
\begin{center}
\includegraphics*[viewport=180 115 550 400,width=80mm,clip]{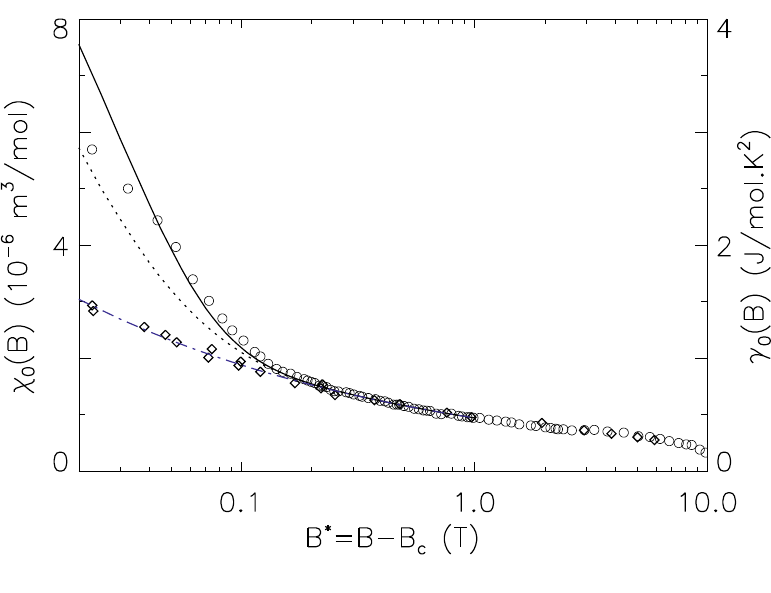}
\end{center}
\caption{The unfreezing of clusters in YbRh$_2$(Si$_{0.95}$Ge$_{0.05}$)$_2$, adapted from Ref. \cite{20}. B$_c$= 0.027 T is the critical field \cite{20}. $\chi_0$ (circles, left axis) and $\gamma_0$ (diamonds, right axis) track each other for B$>$ 0.2 T, and deviate once the clusters unfreeze ($\mu_{cluster}$B$<$ k$_B$T). The dashed curve is a $\sim$1/(B-B$_c$)$^{0.3}$ power law \cite{20}. The dotted curve is the expected response of the collection of AF-clusters at p$_c$ at 0.1 K (using B$^*$ as B to allow for a direct comparison), added on top of the dashed curve. The solid curve is the same as the dotted curve, but with clusters of size 2 given a net moment according to the observed ordering wave vector Q=(0.14,0.14,0) \cite{13}, illustrating that the smallest clusters play a large role in the difference between the $\gamma_0$ and $\chi_0$ curves.} 
\label{fig3}
\end{figure}

At low T and high B/T values, all clusters have aligned with the external magnetic field and their superspin entropy has been shed. As such, clusters do not show up in ac-susceptibility or specific heat measurements and both measurements \cite{20} become temperature independent [$\chi$(B,T)=$\chi_0$(B); c/T= $\gamma_0$(B)], reflecting the response of the conduction electrons partaking in Kondo shielding. It was observed \cite{20} that the Wilson ratio $\chi_0$(B)/$\gamma_0$(B) was B-independent for B$>$ 0.2 T, with strong enhancement at lower B. This enhancement was interpreted \cite{20} as evidence for ferromagnetic fluctuations. However, we show in Fig. \ref{fig3} that the enhancement is due to the unfreezing of the superspins of the clusters, allowing them to react to the ac-field, with increasingly more superspins becoming unfrozen with decreasing B. Thus, the superspins act as ferromagnetic contaminants. Using our computer simulation at p$_c$, we calculate the cluster response (using $\mu_{eff}$= 1.74 $\mu_B$) and find good agreement between the $\chi_0$ and $\gamma_0$ curves. Note however, that the difference between the $\chi_0$ and $\gamma_0$ curves is mostly due to the smaller clusters that far outnumber the clusters with larger superspins. It is unclear to us exactly how well protected the very smallest clusters are from Kondo shielding in an ordered environment; the curves shown in Fig. \ref{fig3} assumed full protection and as such, the good numerical agreement might be somewhat fortuitous. However, the point at B= 0.2 T where  $\chi_0$ and $\gamma_0$ are predicted to start to deviate is largely insensitive to any details concerning the smallest clusters as this point is dictated by the much larger clusters.\\

At the very lowest temperatures, in the presence of a magnetic field, we expect Eq. \ref{acdc} to be a direct test for the presence of clusters. We have estimated the low-T limit for ac- and dc-susceptibility data for YbRh$_2$Si$_2$, and show their comparison according to Eq. \ref{acdc} in Table \ref{table}. While the accuracy of these low-T estimates is not very high bacause of us reading off the values from graphs \cite{20,18}, it is clear that there is a systematic difference between the ac- and dc-susceptibilities. The inferred moment size compares well with the avarage superspin moment determined from our AF percolation simulation. The inferred moment size depends on the fitted value of $\alpha$: we find $\mu_{excess}$= (0.035 $\pm$ 0.015)$\mu_B$ for $\alpha$= 0.3 (exponent used in Fig. \ref{fig2} and in the table), and $\mu_{excess}$= (0.027 $\pm$ 0.018)$\mu_B$ for $\alpha$= 0.32 (not shown in the table). For reference, the expected moment based on our AF-simulation equals 0.026$\mu_B$ (0.0147 uncompensated moments per lattice site times 1.74$\mu_B$ per Yb-ion), with a slighly larger moment expected for departures from AF-ordering. Thus, given the good agreement between the predictions of our cluster scenario and all experimental data, we conclude that clusters with superspins are indeed present close to the QCP in YbRh$_2$Si$_2$, and are responsible for the departures from Fermi-liquid theory.\\
\begin{table}[t]
 {
  \caption{Comparison of low-T ac- and dc-susceptibility limits for YbRh$_2$Si$_2$ using $\alpha$= 0.3.}
 \label{table}
 }
 {
  \begin{tabular}{c c c c}
    \hline
   &   \\[-2pt]
  Field B & $\chi_{dc}$= M/B & $\chi_{ac}/(1-\alpha)$ & [$\chi_{dc}$-$\chi_{ac}/(1-\alpha)$]B \\[3pt]
    (Tesla) & (10$^{-6}$ m$^3$/mol) &  (10$^{-6}$ m$^3$/mol) & ($\mu_B$ per Yb-ion)\\[7pt]
      \hline\\[3pt]
0.1 & 5.8 \cite{18} & 4.0 \cite{20} & 0.026\\[3pt]
0.2 & 3.8 \cite{18} & 2.4 \cite{20} & 0.039\\[3pt]
0.4 & 2.6 \cite{18} & 2.0 \cite{20} & 0.036\\[3pt]
0.8 & 2.1 \cite{18} & 1.5 \cite{20} & 0.075\\[3pt]
0.8 & 1.8 \cite{20} & 1.5 \cite{20} & 0.039\\[3pt]
1 & 1.6 \cite{20} & 1.3 \cite{20} & 0.04\\[3pt]
2 & 1.2 \cite{20} & 1.1 \cite{20} & 0.04\\[3pt]
3 & 1.1 \cite{20} & 1.0 \cite{20} & 0.02\\[3pt]
4 & 1.0 \cite{20} & 0.94\cite{20} & 0.03\\[3pt]
5 & 0.92 \cite{20} & 0.87 \cite{20} & 0.03\\[3pt]
6 & 0.86 \cite{20} & 0.83 \cite{20} & 0.03\\[3pt]
 \hline\\ 
\end{tabular}
  }
\end{table}

Another well-studied quantum critical system is the heavy fermion compound  CeCu$_6$ \cite{23} that is close to a QCP and that can be driven into an ordered phase by substitution of 1:60 Cu ions by Au, or 1:30 by Ag \cite{14}, levels comparable to CeRu$_2$Si$_2$. The approach to ordering in doped  CeCu$_6$ is rather complicated, with the appearance of both incommensurate and commensurate ordering wave vectors \cite{14}. Instead, we argue the presence of clusters based on the results of ac-susceptibility measurements \cite{24} at very low T, deep in the  CeCu$_6$ Kondo-shielded phase.\\

\begin{figure}[t]
\begin{center}
\includegraphics*[viewport=125 105 565 450,width=80mm,clip]{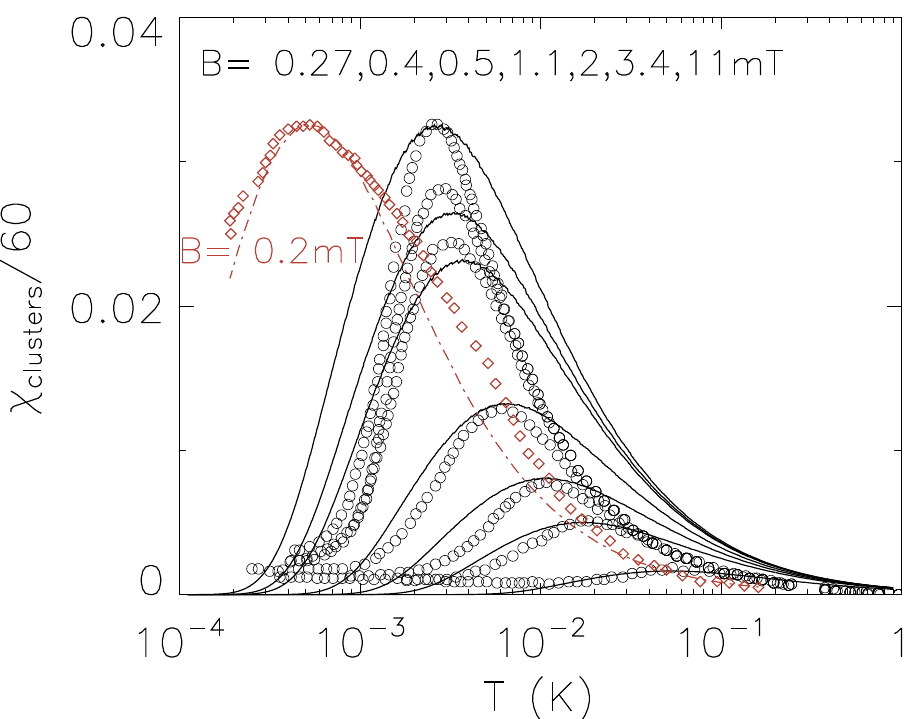}
\end{center}
\caption{(Color online) The measured (circles) \cite{24} and calculated (solid lines) ac-susceptibility of CeCu$_6$ for various fields. Choosing $\mu_{eff}$= 2.1 $\mu_B$, the peaks of the two sets coincide, but the intensity of the cluster distribution has to be reduced in order to obtain agreement in signal strength. The diamonds \cite{25} and dashed curve through the peak \cite{8} are for CeRu$_2$Si$_2$ at B= 0.2 mT and $\mu_{eff}$= 1.2 $\mu_B$ \cite{8}. Note that the peak positions greatly exceed these bare moment values (see text).} 
\label{fig4}
\end{figure}

Ordering fluctuations at the QCP are fairly easy to identify when probing the system as a function of B/T or E/T, but fluctuations along the T=0 K axis should also take place when ZPM fluctuations occasionally drive a Kondo-shielded system to the QCP. In ac-susceptibility measurements we would expect to see a weak echo of the fleeting QCP cluster distribution. Such a signal has indeed been observed in both CeRu$_2$Si$_2$ \cite{25} and in  CeCu$_6$ \cite{24} when probed in the mK-mT range (Fig. \ref{fig4}). We see that most of the signal in  CeCu$_6$ is captured by the response of a cluster distribution at the QCP, provided the strength of the signal is reduced by a factor of ~60. We interpret this as evidence of rare, spontaneous quantum fluctuations from the shielded phase to the QCP. Likely, a similar reduction is also needed in CeRu$_2$Si$_2$ (Fig. \ref{fig4}), however, the data were reported \cite{25} in arbitrary units. Lastly, we note that the observed peak positions (given by $\mu$B= 0.772 k$_B$T) correspond to moment values well in excess of free moments, affirming that superspins must be present in these systems: $\mu$= 6-7 $\mu_B$ for CeCu$_6$;  $\mu$= 4-5 $\mu_B$ for CeRu$_2$Si$_2$;  $\mu$= 4-5 $\mu_B$ for YbRh$_2$Si$_2$. \\

Interestingly, and in contrast to CeCu$_6$ and CeRu$_2$Si$_2$, no out-of-phase component was reported for YbRh$_2$Si$_2$\cite{21}, slightly doped with Ge. A likely reason for this is that YbRh$_2$Si$_2$ was doped to be at the QCP, and as such, a collection of clusters would always be present to react to the probing fields. However, we cannot be absolutely certain that this is a real difference between the systems as it is possible that an out-of-phase component was actually observed in YbRh$_2$Si$_2$: during the experiments such a signal was detected, but it was ascribed to electronic pickup and removed by adjusting the phase factor of the measured signal \cite{jeroen}. The fact that this was possible would also be consistent with the in-phase and out-of-phase components being very similar in the first place, as observed in CeCu$_6$ \cite{24} and CeRu$_2$Si$_2$ \cite{25}.\\

In summary, we have shown that by taking into account the spread in Kondo shielding temperatures arising from doping as well as from zero-point motion that we can account for the most noticeable departures from Fermi-liquid theory, simplifying the inferred phase diagrams in the process. The superspin cluster scenario provides a natural explanation for the inferred ferromagnetic fluctuations, links the entropy and susceptibility and predicts where they differ, and it explains how the moments associated with the positions of the field dependent peaks in ac-susceptibility measurements can exceed those of free ions, and it explains why c/T and ac-susceptibility curves peak at different values (Fig. \ref{fig2}). The cluster scenario allows for a quantitative comparison with experiment over more than two decades in temperare, utilizing only one adjustable parameter that can, in principle and in practice, be ascertained experimentally through high-field neutron scattering experiments as was done in YbRh$_2$Si$_2$ \cite{13}.\\

In addition, the superspin orientation degree of freedom even furnishes the natural low-energy scale \cite{26} required for understanding \cite{li} the observed E/T \cite{27} and H/T \cite{22} scaling in QCP-systems where the response mimics high-energy physics. It also offers an explanation for the observed linear temperature dependence of the resistivity \cite{5,18} in quantum critical systems: since the electronic scattering mechanism tracks the number of fluctuating moments on the infinite cluster, which is in turn being tracked by the entropy, (part of) the resistivity should track the entropy. Thus, when c/T reaches a constant, then the entropy varies linearly with temperature, and so will the resistivity. \\

The zero-point motion induced distribution of Kondo temperatures has not been considered in the literature before, but there can be no reason for ignoring it. Stated differently, ignoring it implies that the adiabatic approximation should fail at low temperatures in quantum critical systems. However, such a putative failure has not been observed. The cluster scenario leaves the Doniach phasediagram \cite{3} intact, with the presence of an infinite cluster separating the two sides of the QCP, and finite clusters present on either side. Also, note that this scenario differs fundamentally from the Griffith phase \cite{28} where rare events dominate the response rather than the collective response of all the clusters.\\

The success of our description also implies that it will be virtually impossible to experimentally test quantum critical theories developed for a uniform heavy-fermion ground state. The zero kelvin fluctuations accompanying a quantum critical point could technically be observed in the response of the infinite cluster. However, at the percolation threshold this cluster is a fractal \cite{9}, and as such, it will have zero weight compared to the finite clusters.

\acknowledgments

We acknowledge stimulating discussions with Alexei Tsvelik, Deepak Singh, and Christian Binek and are most grateful to Jeroen Custers for chasing down old logbooks.

\end{document}